# Comments on testing the violation of the Clausius inequality in nanoscale electric circuits

Elias P. Gyftopoulos
Massachusetts Institute of Technology
77 Massachusetts Avenue, Room 24-111
Cambridge, Massachusetts  02139 USA
and
Michael R. von Spakovsky
Virginia Polytechnic Institute and State University
Blacksburg, Virginia  24061 USA

In Ref. [1], the authors perform calculations about RLC circuits and conclude that the Clausius inequality, one of the foundations of the second law (sic), is broken if the bath temperature is low enough, less than 20 mK.  As a result, they hope that modern experiments may prove that nonthermodynamic energy (sic) flows occur in nature, and break the Clausius inequality đ$Q \leq T$d$S$.  In our view, both the calculations and the hope are not founded on correct thermodynamic and quantum theoretic considerations, and some of the terminology (such as "nonthermodynamic energy") is incomprehensible.

In a new exposition of thermodynamics [2], we start with the concepts of space, time, and inertial mass, and define the terms system, property, and state in a manner that is valid for any paradigm of physics.  Using these concepts and definitions, we express the first law without the concepts of energy, work, and heat.  Some theorems of the first law are the definition and existence of energy $E$ as a property of any system in any state, the conservation of energy, and the energy balance.

Next, we classify states according to their evolutions in time, and introduce the second law (simplified version) as a statement of existence of one and only one stable (thermodynamic) equilibrium state for each set of values of energy $E$, and volume $V$.

Two theorems of both the first and second laws are the existence of entropy as a nonstatistical property of any system, both large and small, in any state, both thermodynamic equilibrium and not, and the entropy balance which avers that any entropy change from one state to another must be accounted for by entropy exchanged with other systems plus a possible but not necessary nonnegative amount generated spontaneously inside the system.

For stable equilibrium states only, other theorems of both laws are that the entropy of any stable equilibrium state must be an analytic function $S\,(E, V)$, and that temperature $T$ and pressure $p$ are defined by the partial derivatives of
$S\,(E, V)$ with respect to $E$ and $V$, respectively.  For the sake of brevity, we omit the dependence of $S$ on the amounts of constituents, and the partial derivatives that define the total potentials.

For two neighboring stable equilibrium states $A_0$ and $A_1$ we can write the differential relation between experimental (tabulated) property data, and the energy and entropy balances.  Specifically, for the experimental data

$$\mathrm{d}S = \frac{1}{T_0}\mathrm{d}E + \frac{p_0}{T_0}\mathrm{d}V \qquad \text{or} \qquad S_1 - S_0 = \frac{1}{T_0}(E_1 - E_0) + \frac{p_0}{T_0}(V_1 - V_0) \qquad (1)$$

Moreover, if the change of state just cited has been brought about by a work interaction (energy only đ$W$), and a heat interaction (energy đ$Q$, and entropy đ$Q/T_R$, where $T_R$ is temperature of the reservoir), then the energy and entropy balances are

$$d E = đQ - đW \quad \text{and} \quad dS = đQ/T_R + \delta S_{irr} \quad (2)$$

By combining Eqs. 1-2 we can find under what conditions the process is reversible or irreversible.

It is important to emphasize that thermodynamics cannot be derived from conventional quantum mechanics because the latter describes only reversible processes that are unitary. But not all reversible processes are unitary, and not all processes are reversible. In addition, a system at temperatures of the order of a few mK cannot be considered as a reservoir both because in that range even minute energy changes result in changes of the temperature of the reservoir, and because relatively large transfers of energy from the reservoir to the system require that the reservoir have negative entropy, a requirement that violates the laws of physics.

We reach identical conclusions in a *nonstatistical* unified quantum theory of mechanics and thermodynamics in Refs. [3-6]. A summary of this scientific revolution is beyond the scope of these brief comments except to say that it is impossible to have "two interacting systems not in a pure state, even though the overall state of the total system may be pure" [1], because then the entropy is simultaneously nonzero and zero (!).